\documentclass[12pt]{article}   
\usepackage{epsfig,times}  % 17 dec. 2001: pbs avec times
\begin{document}
\thispagestyle{empty} 

%\lhead[\fancyplain{}{\sl }]{\fancyplain{}{\sl }}
%\rhead[\fancyplain{}{\sl }]{\fancyplain{}{\sl }}

%%%%%%%% Pour changer les valeurs par defaut pour taille figure,
%%%%%%%% sinon au-dela d'une hauteur de 134 mm = 70% on est rejete a la fin
 \renewcommand{\topfraction}{.99}      
 \renewcommand{\bottomfraction}{.99} 
 \renewcommand{\textfraction}{.0}

%%%%% Definitions

\newcommand{\nc}{\newcommand}

\nc{\qI}[1]{\section{{#1}}}
\nc{\qA}[1]{\subsection{{#1}}}
\nc{\qun}[1]{\subsubsection{{#1}}}
\nc{\qa}[1]{\paragraph{{#1}}}

            % Enumerations
\def\qbu{\hfill \par \hskip 6mm $ \bullet $ \hskip 2mm}
\def\qee#1{\hfill \par \hskip 6mm #1 \hskip 2 mm}

\nc{\qfoot}[1]{\footnote{{#1}}}
\def\qL{\hfill \break}
\def\qpar{\vskip 2mm plus 0.2mm minus 0.2mm}
\def\qtvi{\vrule height 2pt depth 5pt width 0pt}
\def\qth{\vrule height 12pt depth 0pt width 0pt}
\def\qtb{\vrule height 0pt depth 5pt width 0pt}
\def\tvi{\vrule height 12pt depth 5pt width 0pt}

\def\qparr{ \vskip 1.0mm plus 0.2mm minus 0.2mm \hangindent=10mm
\hangafter=1}

                % Decale UN paragraphe
                % Attention! La double accolade est vitale, sinon tout le
                % est decale (cf TEX p.199)
                % On peut aller a la ligne avec \qL=\hfill \break
                % Par contre ne supporte pas les lignes blanches
\def\qdec#1{\par {\leftskip=2cm {#1} \par}}

   %% Defs specifiques
\def\qdpt{\partial_t}
\def\qdpx{\partial_x}
\def\qddpt{\partial^{2}_{t^2}}
\def\qddpx{\partial^{2}_{x^2}}
\def\qn#1{\eqno \hbox{(#1)}}
\def\qds{\displaystyle}
\def\qw{\widetilde}
\def\qmax{\mathop{\rm Max}}   % Petit livre Tex (p.167)
\def\qmin{\mathop{\rm Min}}   % Petit livre Tex (p.167)

%%%%% End of definitions

\def\qci#1{\parindent=0mm \par \small \parshape=1 1cm 15cm  #1 \par
               \normalsize}

\null
% {\large \it To appear in Physica A}
  {\large To appear in the {\it Evolutionary and Institutional Economics Review}}
\vskip 1.5 cm

\centerline{\bf \Large Real estate price peaks: a comparative overview}
%\vskip 5mm
%\centerline{\bf \Large }                                      

\vskip 1cm
\centerline{\bf Bertrand M. Roehner $ ^1 $ }
\vskip 4mm
         
\centerline{\bf Institute for Theoretical and High Energy Physics}
\centerline{\bf University Paris 7 }

\vskip 2cm

{\bf Abstract}\quad 
First, we emphasize that the real estate price peaks which are currently
under way in many industrialized countries (one important exception
is Japan) share many of the characteristics of previous historical
price peaks. In particular, we show that:
(i) In the present episode
real price increases are, at least for now, of the same
order of magnitude
as in previous episodes, typically of the order of 80\% to 100\% .
(ii) Historically, price peaks turned out to be symmetrical with respect to the
peak; soft landing, i.e. an upgoing phase followed by a plateau, has
rarely (if ever) been observed.
(iii) The inflated demand is mainly boosted by investors and high-income
buyers.
(iv) In the present as well as in previous episodes, the main engines
in the upgoing phase have been the ``hot'' markets which developed in
major cities such as London, Los Angeles, New York, Paris, San Francisco
or Sydney. 
In our conclusion, we propose a prediction for real estate prices in the
West of the United States over the period 2005-2011. We also 
point out that investment funds, which already
play a key role in stock markets, have in recent times began to heavily
invest in real estate. In the future, one can expect them to become major
players in property markets worldwide.
The outcome of the present episode will
tell us how quickly this transformation evolves. 
Thus, if the height of the
present peak substantially surpasses the magnitude of
previous ones, one may infer that
investment funds have been able to establish strong 
communication channels between
real estate assets on the one hand and financial assets (e.g. bonds, stocks, 
options) on the other hand.
\qpar

\centerline{July 11, 2005}

%\vskip 8mm
%\centerline{\it Preliminary version, comments are welcome}

\vskip 1cm
Key-words: real estate, speculative trading, property bubble, recession,
debt
\vskip 1cm

1: Bertrand Roehner, LPTHE, University Paris 7, 2 place Jussieu, 
F-75005 Paris, France.
\qL
\phantom{1: }E-mail: roehner@lpthe.jussieu.fr
\qL
\phantom{1: }FAX: 33 1 44 27 79 90

\vfill \eject

\qI{Introduction}

In May 2005, one of my Californian colleagues who closely monitors the 
American real
estate industry wrote to me: ``I think the future is very hard to even
estimate, much less predict, because this market has no historical
precedent that I can identify. For one thing, both originating lenders
and the secondary market seem to have thrown caution to the wind''.
Are we really in unchartered territory? Answering this question will
be one of the main objectives of this paper. Fig. 1 provides a few 
quick preliminary clues. It shows that real estate prices in the West of
the United States experienced four peaks during the past 40 years. The
three previous peaks provide possible guiding lines as to the future of the 
fourth peak that is currently under way. 
%
%%-----------------------------------------------
%%%% Fig.1
  \begin{figure}[htb]
    \centerline{\psfig{width=12cm,figure=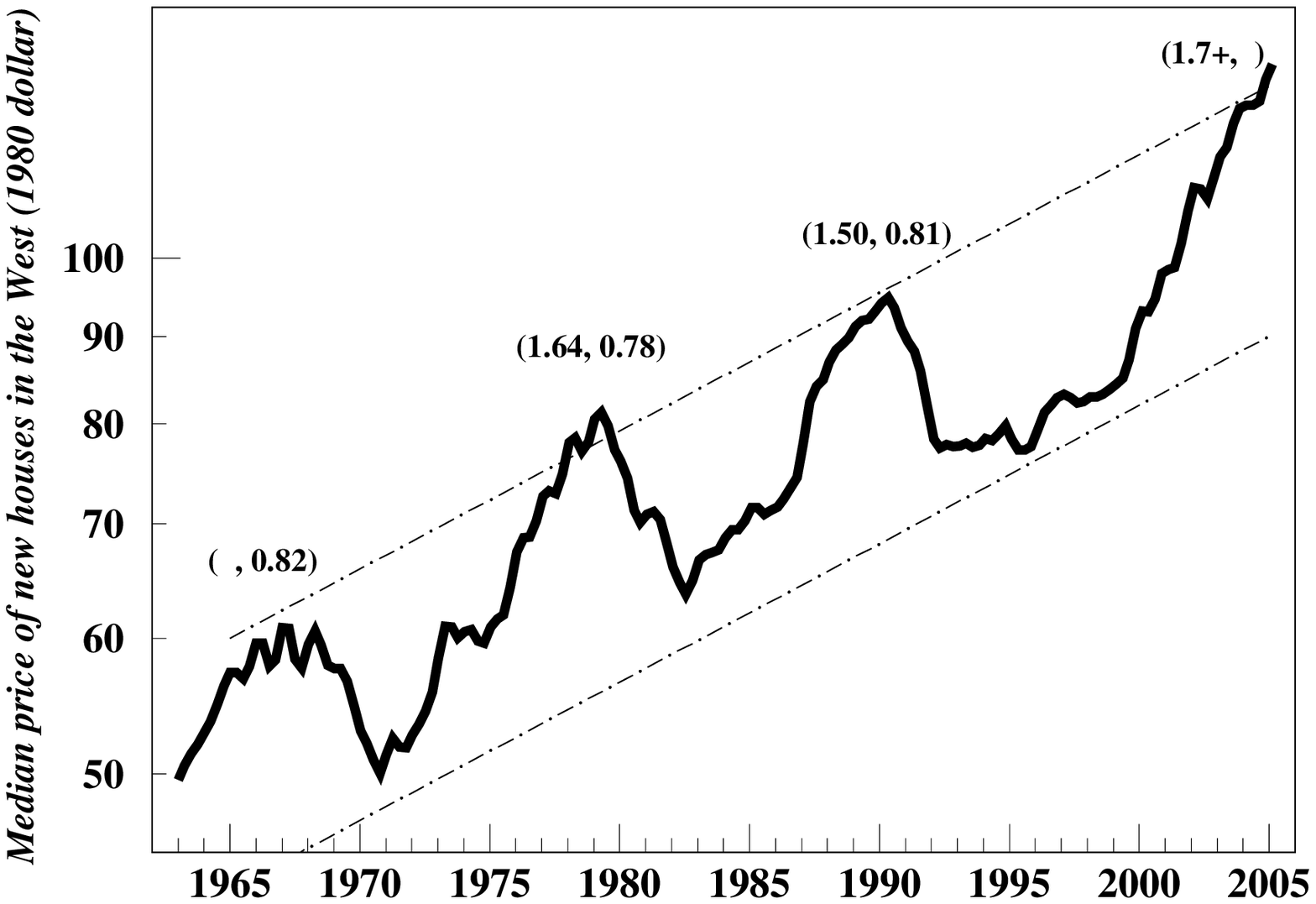}}
    {\bf Fig. 1: Median price of new houses in the West of the United States.} 
{\small The two numbers above the peaks give the amplitude 
of the peak (the ratio of
peak price to initial price) and the amplitude of the fall in the downgoing
phase (ratio of trough price to peak price). The notation 1.7+ shows that
the current peak is still in its upgoing phase and will have an amplitude
larger than 1.7.
The whole curve has been smoothed
using a three-year centered moving average.}
{\small \it Source: U.S. Bureau of Census}
 \end{figure}
%% --------------------------------------------------
%
Naturally, there is no 
absolute certitude 
that the present peak will unfold as the previous ones; there are mainly
two new factors (i) The present peak has a bigger amplitude and duration
than the previous ones; in itself this would probably not preclude a repetition
of the previous scenarios (ii) Investment funds (in which we include pension
funds, equity funds, hedge funds) have taken a much greater part in the
present episode than in previous ones. This explains the exceptional
size of the peak but, as these institutions can mobilize much larger
amounts of capital than the real estate companies which operated in 
previous episodes, they may be able to stage a softer landing. 
We come back to this question in the conclusion.
\qpar

The paper focuses on three issues.
\qbu It emphasizes that real estate price peaks occurred repeatedly in the
19th and 20th centuries.
\qbu It examines whether during peak episodes, 
prices are driven by demography or by speculative trading.
\qbu It shows that the ``engines'' of peak price episodes are
the ``hot'' markets of big cities.
\qpar

From the point of view of economic theory, the second question is crucial
because it tells us which variables a model should contain. If prices were
driven by demography the model would have to include a large number of
exogenous variables such as for instance population change, revenue
increase, price of construction (salaries, materials, interest rates), etc.
On the contrary, if speculative trading is the main moving force, price peaks
may be described in the framework of self organized criticality introduced
by late Per Bak. The task of developing such a model will be left to 
a subsequent paper. 
\qpar

From a macroeconomic perspective, real estate price peaks are of 
great importance. Let us recall that the recession experienced by
the Californian economy between 1991 and 1995 had its origin in the
real estate market crash of 1990 (for more details see Roehner 2002, p. 115-117).
On an even larger scale, the property crash in Japan during the 1990s affected
adversely the Japanese economy. It lead to a disappearance of wealth that
amounted to 1.6 quadrillion yen that is to say twice the country's GDP.
Instead of using about 10\% of their salaries to service their loans, 
Japanese households had to devote more than 20\% of their salaries
to the repayment of their debt (Straits Times, November 9, 2002).
At the time of writing, the price of land in Japan is still declining (Wall
Street Journal, July 11 2005): since mid-1991 it has been divided by
2.1.

\qI{Historical examples of real estate price peaks}
In order to give a microeconomic underpinning
to real estate price surges,
let us briefly recall an episode which took place during the rush to the
West in the late 19th century.
As ticket prices to Southern California dropped substantially, emigration 
soared. More than 41,000 people came to San Diego between 1885 and 1888.
By the peak of the building boom in 1888, at least 10 brickyards employing 
over 500 workers were operating in San Diego. In late 1888, San Diego's 
real estate market began to deflate. One of the largest brickyards, the 
Park Brick Yard Company, went out of business. By 1889, people were
leaving in droves, driving San Diego's population down to less than
17,000. A similar real estate boom occurred in Los Angeles; it came to its
end in 1889, that is to say a few months after the one in San Diego. 
\qpar

Were the market collapses in Los Angeles and San Diego brought about
by a nationwide economic recession? The answer is no. As a matter of
fact in 1888-1889, with an unemployment rate as low as 4\%,
the state of the economy was good. It is only in subsequent years, with
the onset of the depression of 1893-1894 that unemployment rates picked
up, reaching levels of 10\% in 1893 and 15\% in 1894 (see the data
provided by David O. Whitten in the article about the Depression of
1893 in the eh.net encyclopedia). 
\qpar

Did the crash of the property market in California have an impact on
the economy nationwide? Again, the answer is no. As we have just seen, the 
recession occurred 6 years later and can of course not be attributed 
to the property crash. How can one explain that it had no
sizable impact on the economy? In 1890, California had a population
of 1.2 million which represented less than 2\% of the US population.
Moreover, the property crash in the West was {\it not} paralleled by a
similar crash in the East. This is made clear by Fig. 2a,
%%-----------------------------------------------
%%%% Fig.2a
  \begin{figure}[tb]
    \centerline{\psfig{width=12cm,figure=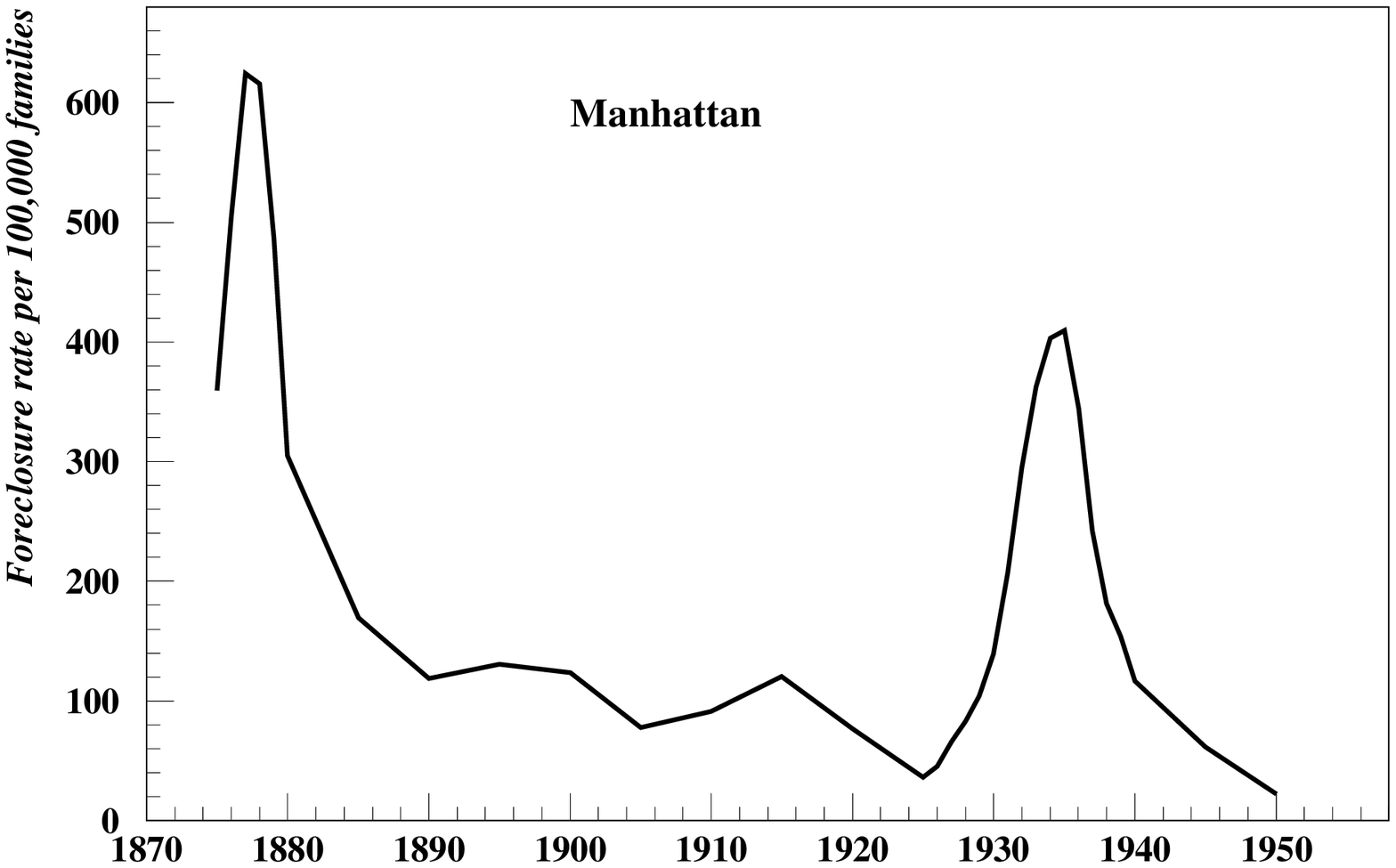}}
    {\bf Fig. 2a: Foreclosures in Manhattan (New York).} 
{\small When a house owner becomes unable to repay mortgage
installments a foreclosure sale takes place. The two peaks correspond
to the recession of 1878 and to the Great Depression respectively.}
{\small \it Sources: Wenzlick (1972), Kaiser (1997)}
 \end{figure}
%% --------------------------------------------------
which shows
the mortgage foreclosure
rate in Manhattan. Obviously there was no
surge in foreclosures in 1888-1889%
\qfoot{More surprisingly, there was only a small increase in foreclosure
numbers during the recession
of 1893-1894 in marked contrast with what happened during the
Great Depression.}%
.
\qpar

What can one learn from this example?
\qbu Firstly, real estate booms seem to have their own dynamic. 
Property crashes occur when prices have reached unsustainable levels
without being necessarily triggered by a nationwide recession or by a
jump in interest rates%
\qfoot{Short-term interest rates of commercial paper were in fact at a
(temporary) low in 1888-1889 as can be seen from the following time series
(Homer and Sylla, 1996, p. 320). 1887: 5.73\%, 1888: 4.91\%, 
1889:4.85\%, 1890: 5.62\%, 1891: 5.46\%, 1892: 4.10\%, 1893: 6.78\%, 
1894: 3.04\%.}%
.
\qbu Secondly, property markets in the North-East and in the West
were quite disconnected. It is only in the last decades of the 20th century
that they have become more correlated. 
\qpar

Other historical examples of real estate price peaks are given in
Fig. 2b, 3 and 4. Fig. 2b is of interest because it emphasizes the similarity 
in the shape of two peaks which occurred in different time periods and
in distant countries. In the same way, Fig. 3 points out the close parallelism
between the peaks of 1889 and 1929: they have the
same amplitude (amplitude 
of a peak being defined as the ratio (peak price) / (initial price)) and
almost the same duration. Fig. 4 shows the 1985-1995 price peak in
Britain (more will be said about this case in section 4). 

%%-----------------------------------------------
%%%% Fig.2b
  \begin{figure}[tb]
    \centerline{\psfig{width=11cm,figure=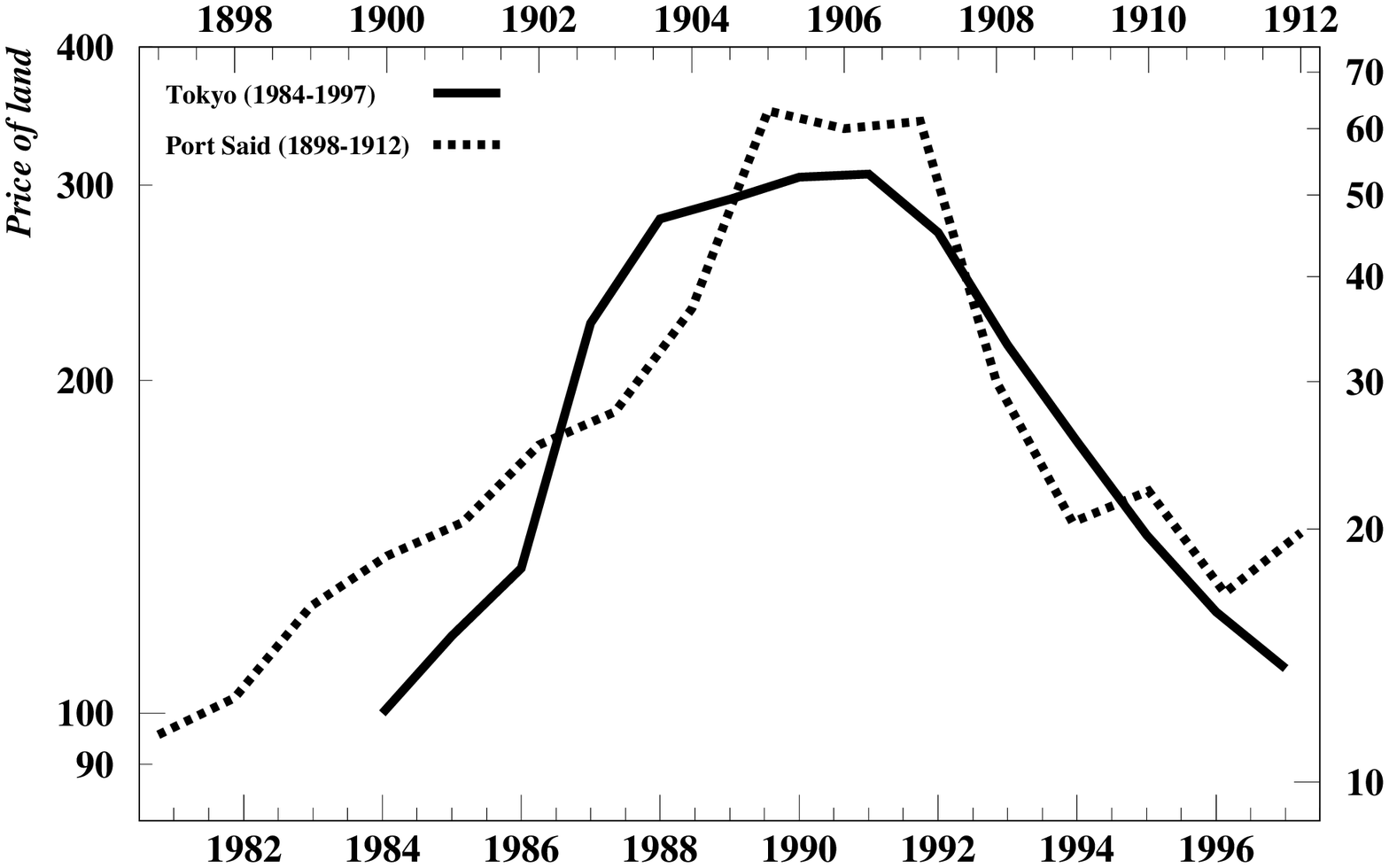}}
    {\bf Fig. 2b: Price peaks for land.} 
{\small Solid line: annual (nominal) prices
of commercial land in the Tokyo area (left-hand side and bottom scales).
Dashed line: annual price of land in Port Said at the entrance to the
Suez Canal, expressed in French francs per square meter (right-hand side
and top scales). The Suez canal was opened in 1869.
Although these episodes are almost one century apart 
they are fairly similar in the sense that in both cases 
the property crash came along with a stock market crisis. This 
association between stock and real estate crisis is fairly common:
the simultaneous
stock and property crashes in Paris in 1882 and in 1931 are 
two other instances.}
{\it \small Sources: Bourgeois (1913), Grison (1965),
Financial Times (17 October 1997)}
 \end{figure}
%% --------------------------------------------------
%
%%-----------------------------------------------
%%%% Fig.3
  \begin{figure}[tb]
    \centerline{\psfig{width=11cm,figure=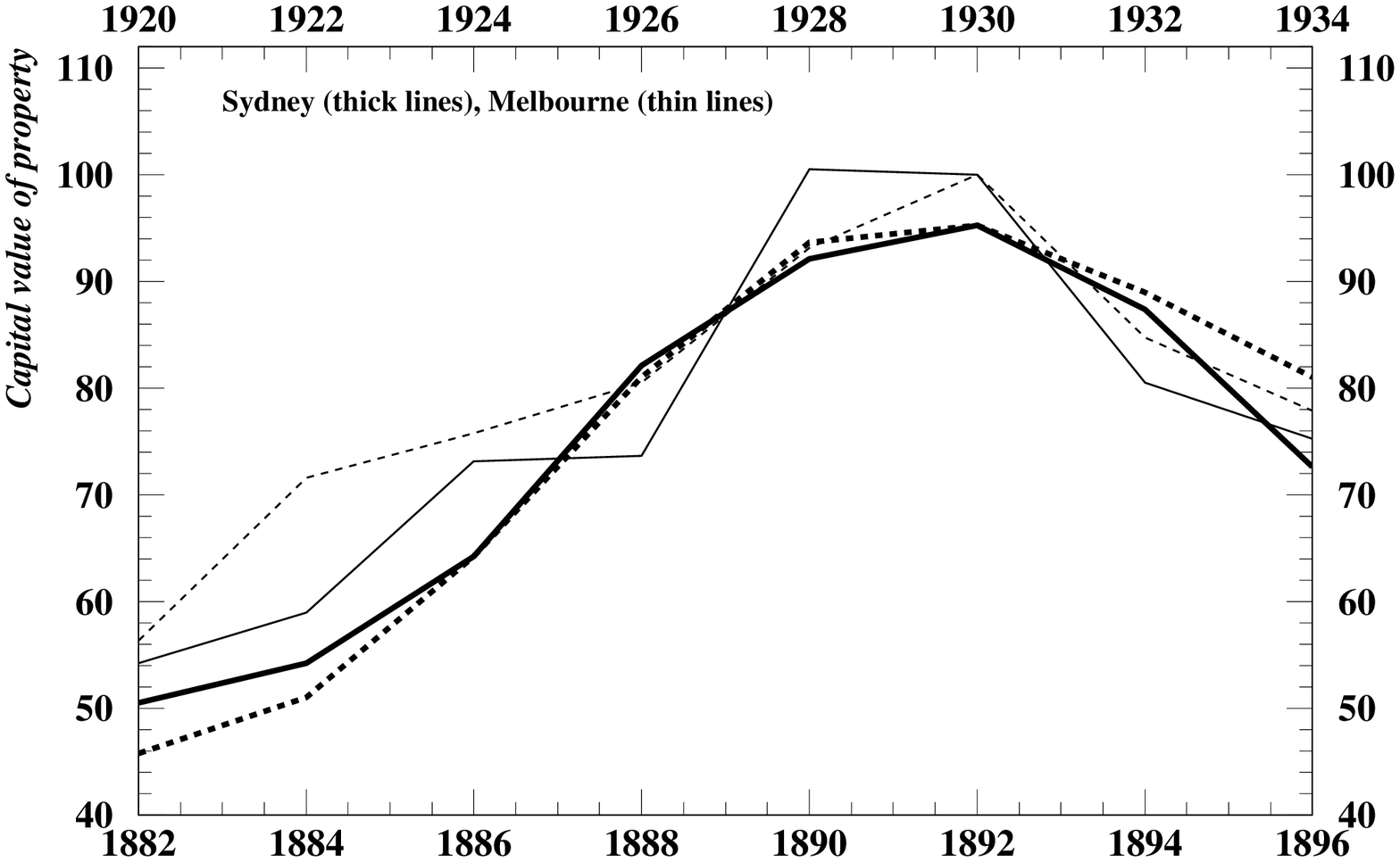}}
    {\bf Fig. 3: Property valuation in Sydney and Melbourne (Australia).} 
{\small Solid lines: 1882-1896 (left-hand side and bottom scales),
dashed lines: 1920-1934 (right-hand side and top scales). 
Note the parallelism between the 1890 and 1930 episodes.
It must be observed that
accurate time series
of property price data are not available; the graphs rather show estimates
of property value.}
{\small \it Sources: Fisher and Kent (1999)}
 \end{figure}
%% --------------------------------------------------
%
%%-----------------------------------------------
%%%% Fig. 4
  \begin{figure}[tb]
    \centerline{\psfig{width=12cm,figure=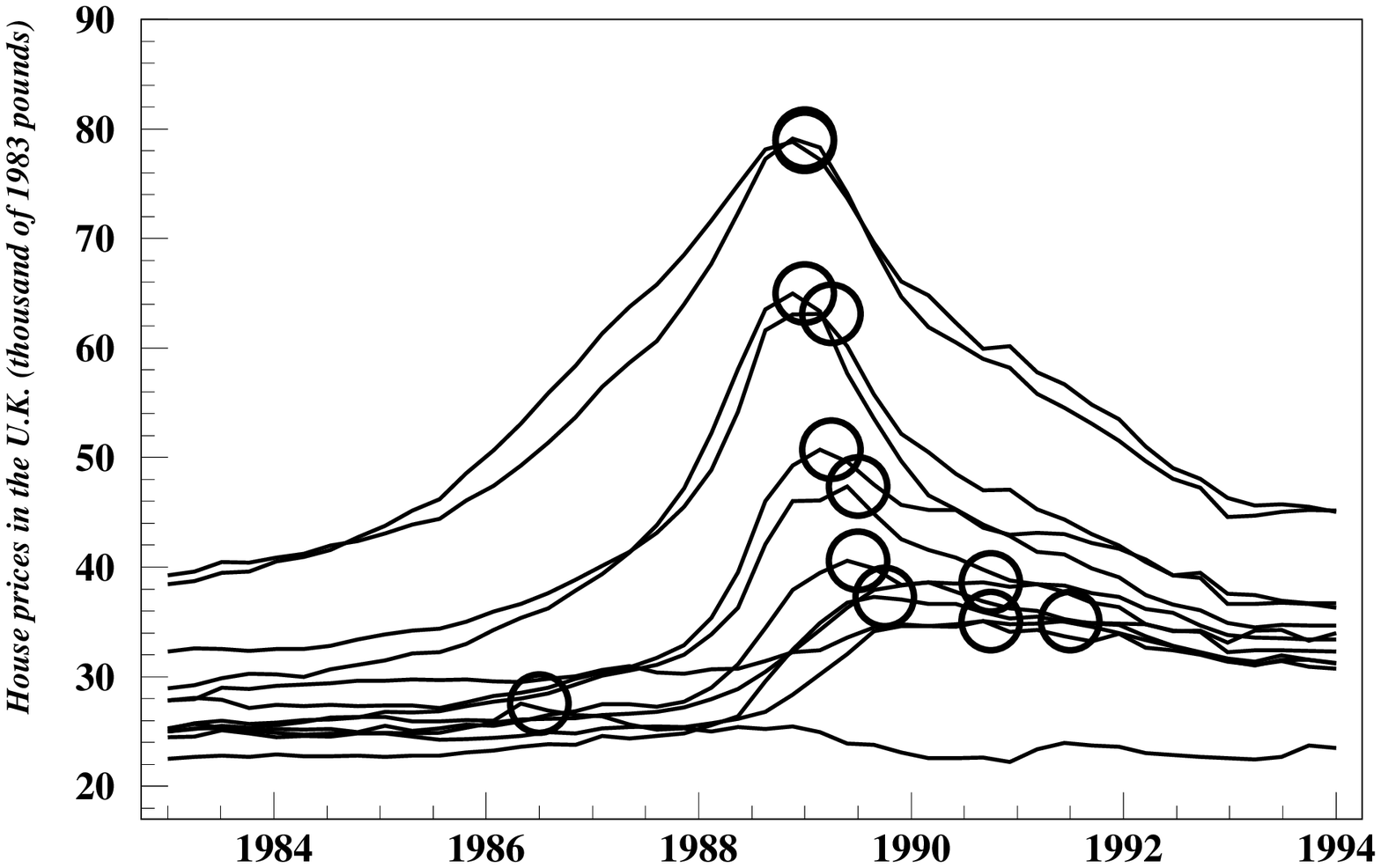}}
    {\bf Fig. 4: House prices in Britain during the speculative
episode of 1984-1994.} 
{\small Each curve represents the price in one of the 12 regions
composing the United Kingdom. The two
highest curves correspond to ``Greater London'' and ``South East''. 
The circles indicate the positions of the maxima; the more the region
is distant from London, the later it reached its maximum.
Note that the obvious outlier corresponding to the 
maximum of the lowest curve (Northern Ireland)  should not be
taken seriously
for this curve, in fact, has no real maximum.}
{\small \it Source: British property data are collected
and published by the Halifax group (West Yorkshire, England). I am most
grateful to the people at Halifax for their kind assistance.}
 \end{figure}
%% --------------------------------------------------
\qpar

These various examples along with additional evidence given
in Roehner (2004, p. 112-114) suggest the following rules of thumb.
\qee{1)} Roughly speaking price peaks are almost symmetrical with 
respect to their maximum, which means that the rising and falling
phases have approximately the same duration. 
\qee{2)} The total duration of a peak is about 12-14 years which means that the 
upgoing and downgoing phases last about 6-7 years%
\qfoot{A closer look shows that usually the falling phase
is somewhat shorter than the rising phase, therefore a more accurate
rule would be: rising phase $ \sim 6-7 $ years, plateau in the vicinity
of the peak $ \sim 2 $ years, falling phase $ \sim 4-5 $ years.}%
.
\qee{3)} Usually the amplitude of the peak is less than 3 because price
increases are restrained by income levels%
\qfoot{When a property boom is primarily driven by commercial capital
as was the case in San Diego or Port Said, the amplitude can be
larger than 3.}
.

\qI{Are price peaks driven by demography or by speculative trading?}

By the expression 
``driven by demography'' we mean driven by population and average
income increases. Driven by speculation means that a sizable part of
the houses are bought by investors who plan to sell them one or 
two years later; speculative trading
also includes transactions in properties which are still in their
design stage (sometimes referred to as ``off-the-plan transactions'')
which are sold one or two years {\it before}
actual completion. 
\qpar

Whether or not a property boom is driven by speculation is always a 
hotly-debated question because of its obvious implications. If the boom
is driven by demography, one may expect that the period of price increase
will be followed by a plateau without any substantial price fall. On the contrary,
if prices are pushed up by speculation, one should expect a price pattern
similar to those described in section 1. Needless to say, such a price pattern
may scare away potential buyers, which is why real estate agents may be tempted
to present it as a remote and unlikely perspective%
\qfoot{The fact that the views of
real estate analysts may be affected by conflicting interests
does not facilitate the emergence of
a clear understanding. This can 
even restrict data availability; thus, in the period 1999-2003
it was very easy to find (free)
Australian real estate price series on the Internet, but after the downturn many
real estate institutes restricted access to subscribers. Similarly,
after the downturn, in spite of prices falling at annual rates of 8\%-10\%,
real estate experts would continue to use such euphemisms
as ``the flat market in Sydney'' or ``the market has cooled down''. }%
.
Here we will restrict ourselves to
listing a number of criteria which should help us to come up with a definite
conclusion.
\qee{1)} Are the increases in property prices much faster than GDP
growth? 
\qee{2)} Are prices of expensive houses rising
faster than prices of more affordable houses?
\qee{3)} Are stock prices of real estate companies climbing to towering
levels?
\qee{4)} Does the proportion of transactions carried out by investors
grow along with the level of prices?
\qpar

As an illustration of the first criterion one can mention the case of
Hong Kong between 1992 and 1998 (see Fig. 8). The average price of
apartments jumped from US\$ 2,100 to US\$ 8,600 which represents an
annual growth rate of 41\%. Obviously this is completely disconnected
from GDP growth.
\qpar

The second criterion is closely related to the price multiplier effect which
is explained in the next section. 
It has been shown elsewhere 
(Roehner 2000, 2001 chapter 6, 2002 chapter 7, Maslov et al. 2003)
that this phenomenon, referred to as the price multiplier effect,
can be observed in all kinds of speculative price peaks whether
in property, collectible stamps, rare books or stocks. In a sense, it
can be seen as a signature of speculative price peaks. A related consequence
is described in Fig. 5: it shows that as price climb sales tend to
concentrate on the most expensive market segments; this effect is particularly
strong in the North East of the United States.
\qpar

Fig. 5 and 6 provide two elements which should permit to answer the
question in criteria 2 and 3 for the period 1995-2005.
%
%%-----------------------------------------------
%%%% Fig.5
  \begin{figure}[tb]
    \centerline{\psfig{width=11cm,figure=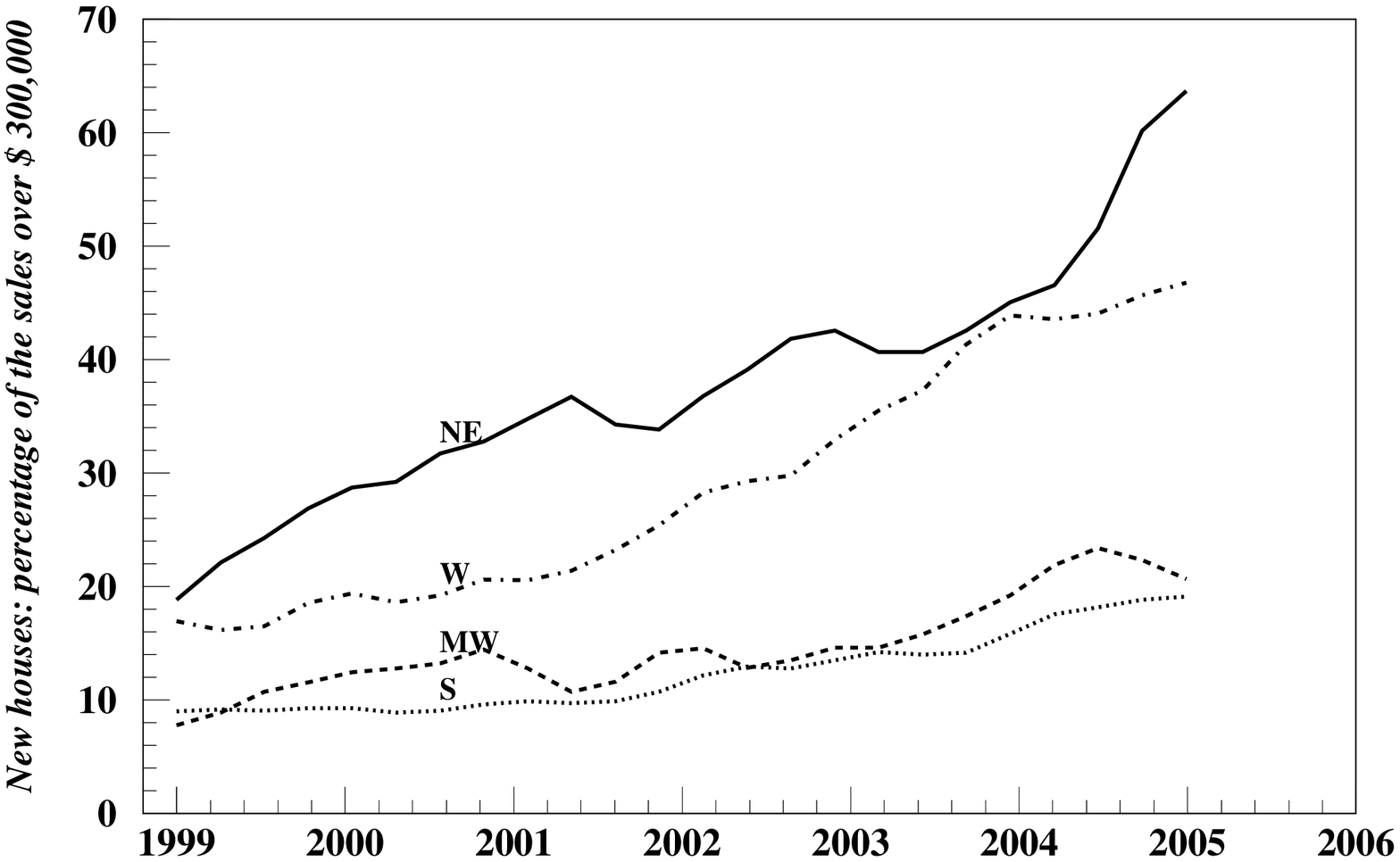}}
    {\bf Fig. 5: Percentage of sales over \$ 300,000.} 
{\small If demand were driven solely by demographic factors and
revenue increase, there would be no reason for the huge increase
in the percentage of expensive home sales. Note that there is a similar
increase in the percentage of transactions over \$ 1 million: in 
2002 they represented 3.0\% of the transactions in California whereas
in 2004 they represented 5.0\%. According to DataQuick,
Ross in Marin County and Rancho 
Santa Fe in San Diego county were communities where virtually 
all home sales were in the million-dollar category.}
{\small \it Sources: DataQuick Real Estate (DQNews.com); US Bureau
of the Census: New residential sales (New houses sold by sales price).}
 \end{figure}
%% --------------------------------------------------
%
%
%%-----------------------------------------------
%%%% Fig.6
  \begin{figure}[tb]
    \centerline{\psfig{width=11cm,figure=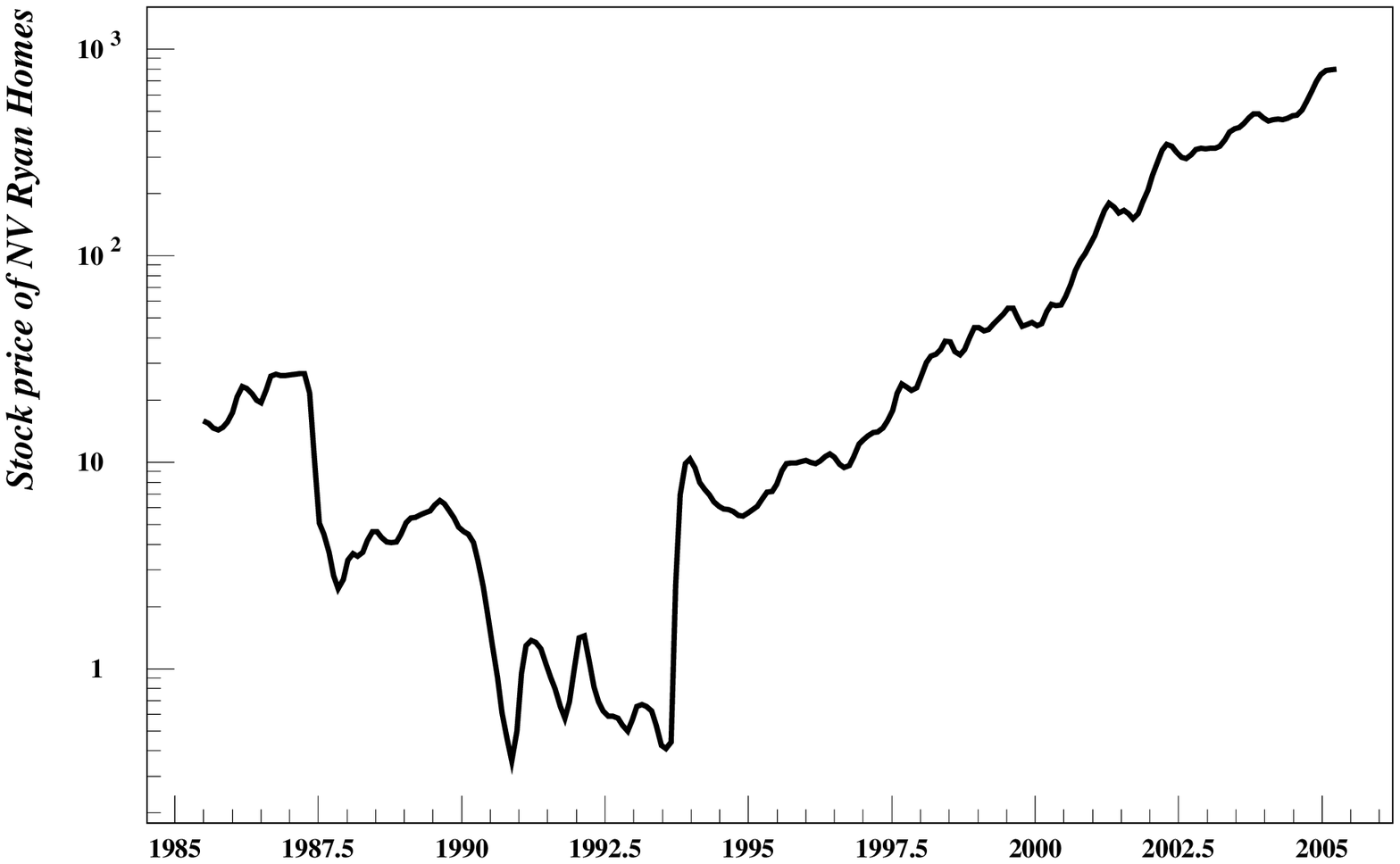}}
    {\bf Fig. 6: Stock price of NV Ryan Homes (AMEX: NVR).} 
{\small NVR operates in home building and mortgage banking in
11 states mainly in the eastern part of the United States. In the
10-year time interval between 1995 and 2005, its stock price was
multiplied by one hundred. In late March 2005 its market capitalization
was one third of the capitalization of General Motors.
The price fluctuations in the 1985-1994 period
to some extent reflect the property boom which, at least in the North-East,
peaked around 1988 and bottomed around 1994.}
{\small \it Source: http:// finance. yahoo.com}
 \end{figure}
%% --------------------------------------------------
%

\qpar

The fourth criterion would probably be the most effective; unfortunately,
one has only fairly scattered statistical data; for instance, one knows that
between 2000 and 2004
in the United States, the share of purchases made by investors increased
from 5.8\% to 8.5\% (Chattanooga Times Free Press, March 20, 2005).
Naturally, as one could expect this percentage is higher in ``hot'' markets
where price increase faster. Thus, in the first quarter of 2005, they represented
on average 11\% in 30 metro-areas; in places like Phoenix, Las Vegas or in
many parts of California, transactions by investors represent 15\% 
(USA TODAY, June 23, 2005). 
\qpar

Finally, the very fact that during a price peak episode, prices are often
rising simultaneously in several industrialized countries irrespective
of local conditions shows that a common driving force is at work in all
these markets. At the time of writing (July 2005) prices are still rising
in California as well as in 
Paris, but they tend to flatten in Britain whereas in Melbourne or Sydney they
have been falling for about one year.

\qI{What fuels price peaks? The key role of big cities}
There are (at  least) two different types of property booms, the commercial
type and the residential type. Examples of the commercial type are provided
by San Diego and Port Said, as discussed in section 1. Needless to say,
neither San Diego nor Port Said were big cities in the 1880s. In these booms,
apart from residences, 
the rush also included hotels, banks, shops, public buildings and so forth.
In contrast, the boom of 1985-1995 in Britain concerned only residential
property. What we say in this section applies only to the second type of
real estate booms. 
\qpar

The major role played by big cities is illustrated by the fact that in the
United States there have been price peaks of an amplitude greater
than two almost only in the North-East (Boston, New York) and in
California (Los Angeles, San Francisco) and marginally in the Chicago
Metropolitan Area. Fig. 4 further explains how price increases spread from
big cities to neighboring areas. The highest curve corresponds to London;
as can be seen the time lag between the maximum in London and
the maxima in northern counties is comprised between one and two years.
\qpar

This observation has interesting practical implications. It means
that the market downturn can first be observed
in the hottest places. For instance,
during the third and fourth quarter of 2004, prices in London have been
falling, whereas they were still increasing in the rest of Britain (albeit
at a slower rate than earlier). This seems to suggest that
by July 2005 (at time of writing)
Britain already was in the downward phase of the price peak. 
The same conclusion
holds for Australia, where Sydney (which is the hottest market)
has seen declining real estate prices since 2004.
\qpar

The fact that price increases are positively correlated with prices at
the beginning of the price peak is illustrated in Fig. 7 in the case of
the Western region of the United States. Thus, the average 
price in Las Vegas,
which was \$ 120,000 in 1995 had been multiplied by 1.2 in 2002,
whereas the average price in San Francisco, namely \$ 280,000
in 1995, had been multiplied by 1.8 in 2002. Additional evidence about
the price multiplier effect can be found in Roehner (2000),
Roehner (2001, chapter 6), Maslov et al. (2003), and Roehner (2004).
%%-----------------------------------------------
%%%% Fig.7
  \begin{figure}[htb]
    \centerline{\psfig{width=14cm,figure=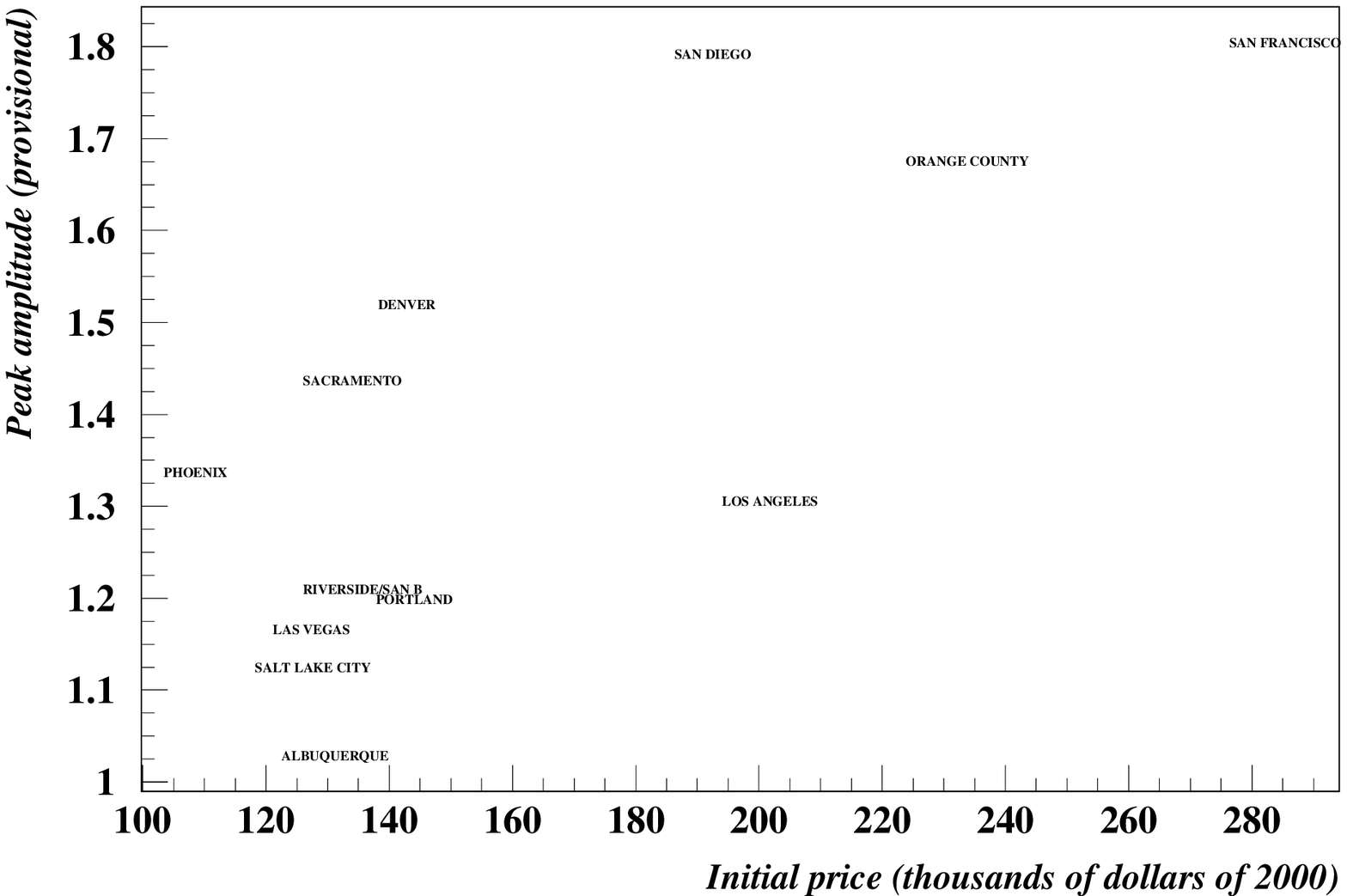}}
    {\bf Fig. 7: Correlation between initial housing prices
and peak amplitudes, West of the United States, 1995-2002.} 
{\small Peak amplitude is here defined as the ratio: (price in 2002)/(price in 1995);
in 2002 the peak was not yet reached which means that these are 
in fact provisional
peak amplitudes; the correlation between initial prices and
(provisional) peak amplitudes is equal to 0.75.}
{\small \it Source: Websites of the California Association of
Realtors and of the National Association of Realtors.}
 \end{figure}
%% --------------------------------------------------

\qpar

In section 2 we observed that prices continue to climb until they
reach an ``unsustainable'' level. How can this level be
identified? Figure 8 provides a clue. It shows
that during a property boom, prices increase faster than rents. 
Therefore, there is a moment when the yield%
\qfoot{The annual rent represents the income brought in by an apartment;
therefore the ratio: (annual rent) / (price of the apartment) represents
the yield of the apartment in the same way as the ratio: (annual coupons) /
(price of the bond) represents the yield of a bond. Similarly, the ratio:
(annual dividends) / (stock price) represents the yield of a stock; for stocks
it is more common to use the price-earnings 
ratio: (stock price) / (annual dividends) which is the inverse of the yield}%
of apartments becomes lower than the average yield of financial
markets. At that point the price increase becomes the only real source
of profit; this is an unstable situation in the sense that any downward price
fluctuation may generate a spate of selling and trigger a market wide
downturn.
\qpar

%%-----------------------------------------------
%%%% Fig. 8
  \begin{figure}[htb]
    \centerline{\psfig{width=12cm,figure=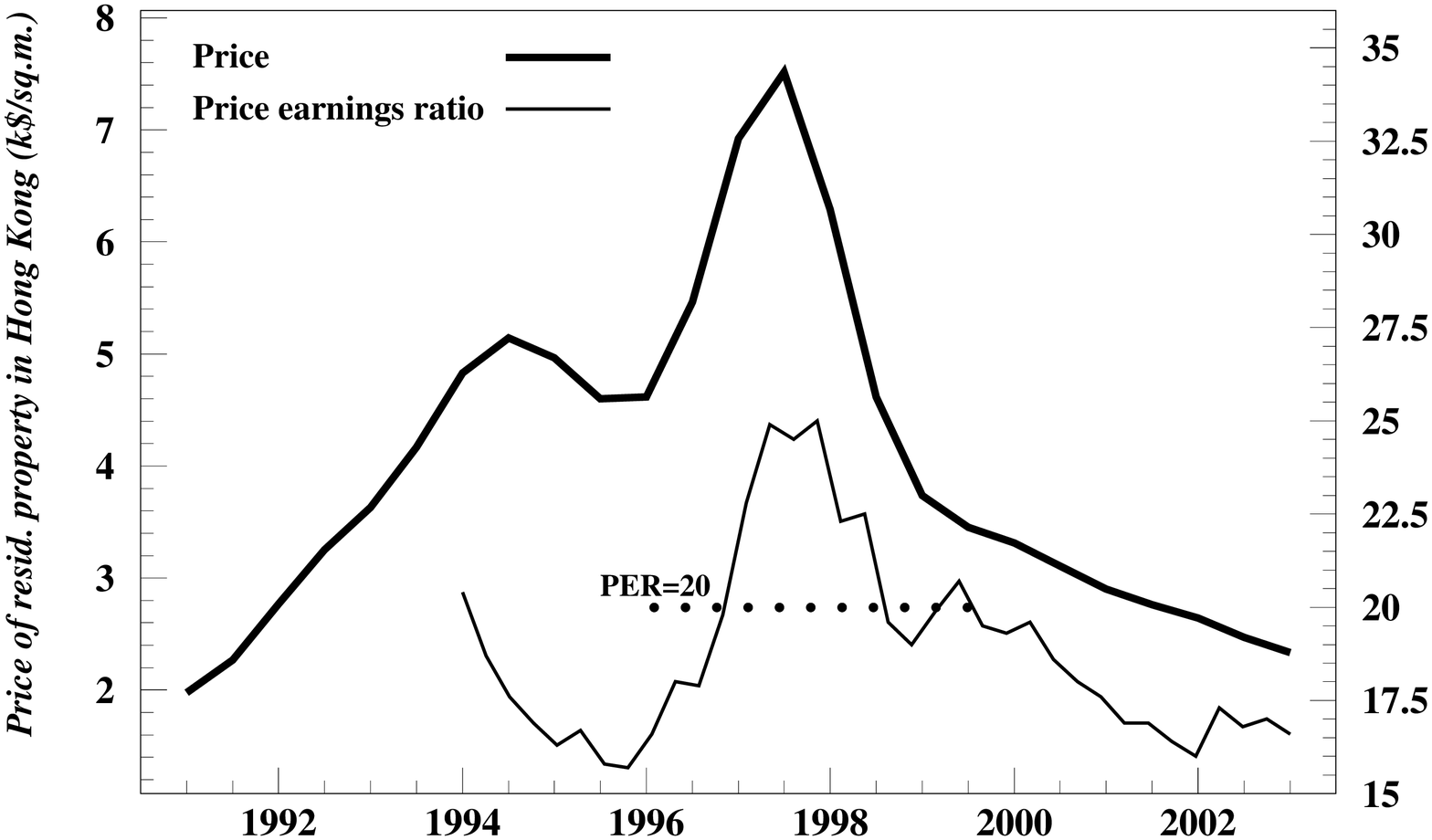}}
    {\bf Fig. 8: Price and price-rent ratio in Hong Kong during
the speculative episode of 1992-2002.} 
{\small The prices are expressed in $ 10^3 $ US\$ per 
square meter.
When seen from the owner's (instead of tenant's) perspective
the price-rent ratio should rather be called
a price earnings ratio; this is why we used this expression
in the graph; it has the additional advantage of establishing
a clear link with the price-earnings ratio of stocks.}
{\small \it Source: Website of the Rating and Valuation 
Department of the Hong Kong Government}.
 \end{figure}
%% --------------------------------------------------

{\bf What is the role of interest rates?}\quad As already mentioned,
real estate price peaks have their own dynamic. While declining interest
rates may boost prices and transaction volumes in the upgoing phase, 
they cannot stop a
market from falling once the downturn has occurred.
This was demonstrated in the United States
in 1992-1994, a period of declining interest rates {\it and} of falling
real estate prices in California.
More recently in 2004, the fall in real 
estate prices in Australia continued 
in spite of declining interest rates:
in July 2002, at the height of the real estate boom,
the yield (i.e. interest rate) of the 10-year (Australian) Treasury bond 
was 5.86\%; 
in November 2004 it was down to 5.22\% 
but this did not prevent the downward spiral
of property prices in Sydney to continue.
\vskip 3mm

\qI{Conclusion}

In this conclusion we first present a projection for prices
in the West of the United States over the period 2005-2011.
Then, we explain why it may be possible to offer
predictions for real estate prices
whereas making predictions for stock markets is much more difficult 
if not altogether impossible. 
\qpar

Fig. 9 provides a synoptic view of three of the peaks
represented in Fig. 1 along with the 2005-2011 projection.
In previous papers we described speculative price peaks by
a function of the form:
$$ p(t)=p_2 \exp \left[ -\left| { t-t_2 \over \tau }  \right| ^{\alpha} \right] $$

where $ p_2, t_2 $ denote the peak-price and peak-time respectively; 
$ \alpha $ and $ \tau $ are two adjustable parameters.
In the present case it turns out that the exponents $ \alpha $ are almost equal
to 1, namely 1970-1982: $ \alpha =0.99 $, 1982-1992: $ \alpha =1.06 $; 
the parameters
$ \tau $ turn out to be almost the same as well, namely of the order 
of 13.5 quarters (i.e. 3.3 years). 
As a result, one is encouraged to model the downgoing
path of the current episode by the same parameters. 
This leads to the dotted line projection in Fig. 9%
\qfoot{A prediction about the moment of the downturn (which we did not
try to predict here) can be found in a recent paper by Zhou and Sornette
2005; another related (and groundbreaking) contribution to this debate
is provided by Taisei Kaizoji (2005).}%
.
Needless to say,
this projection rests on the assumption that there is no fundamental change
with respect to the two previous episodes. In particular, we assume that
in spite of their expanding assets, investment funds will not be able to rule
property markets in coming years
to the same extent as they are able 
to direct stock markets.
\qpar

%%-----------------------------------------------
%%%% Fig. 9
  \begin{figure}[htb]
    \centerline{\psfig{width=12cm,figure=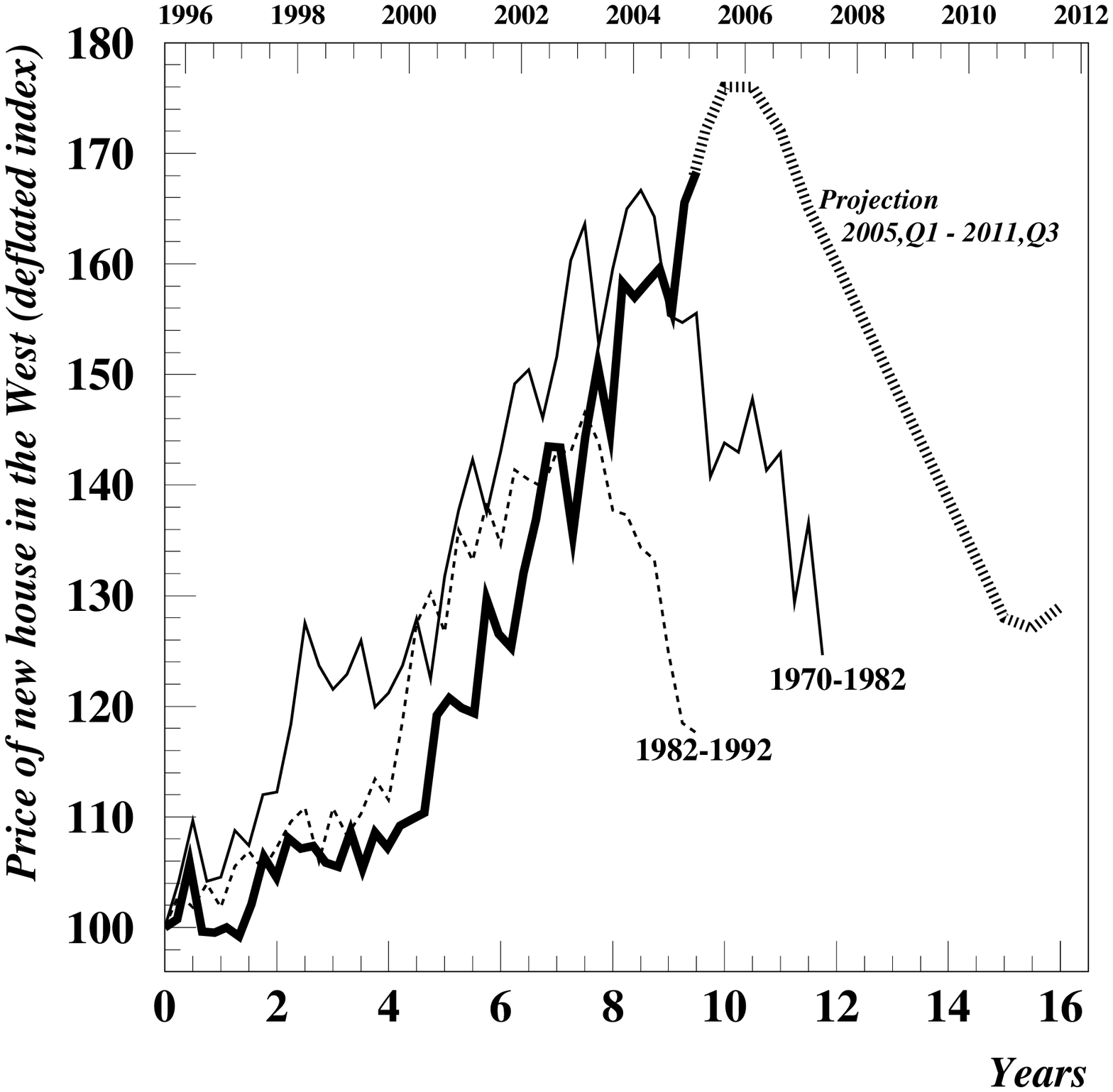}}
    {\bf Fig. 9: Projection for the price of new houses in the West of the
United States: 2005-2011.} 
{\small The figure summarizes three of the
four episodes shown in Fig. 1. The two
downgoing phases are characterized by exponential price falls
with rates of the order of $ -6.5\% $. The projection was modeled on
the same pattern. Note that the precise
moment of the downturn cannot be forecast in the same way because it
depends on exogenous factors such as interest rate levels or exchange rates.
However, after the downturn 
one can expect a subsequent 5-year period characterized by a
downward trend with a rate averaging about $ -6\% $ per year.}
{\small \it }.
 \end{figure}
%% --------------------------------------------------

Apart from their own specific interest, 
speculative episodes in property markets
are also of great value because they are similar to, but simpler than,
speculative episodes in stock markets. The first point, the similarity
of price peaks in property versus stock has been briefly summarized
above (more details can be found in Roehner 2004a). The fact that property
markets are ``simpler'' than stock markets can be attributed to the
following circumstances. 
(i) Transactions take much longer in real estate
than in stocks, typically one or two months compared to one or two minutes;
as a result property prices are subject to only low frequency shocks whereas
stock prices are subject to shocks whose frequency
spectrum extends over several orders of magnitude (from 1/minute to 1/year).
(ii) Most of the financial instruments available on stock markets (such as
for instance options, futures, convertible bonds) do not (yet) exist in
property markets.
(iii) As shown in a former paper (Roehner 2004b), the strategy of 
big investment funds
have  a determinant impact on the price of stocks%
\qfoot{One can recall in this respect the action of Jeffrey Vinik, the manager
of Fidelity Magellan, a fund of parent company Fidelity Investment
whose assets represent about 10\% of the American GDP. In late 1995,
concerned by
earning growth problems
in the technology sector, he cut the fund's investment in
this sector from 43\% to 24\%. Although perhaps sound in itself,
Vinik's strategy came too soon and misfired. Disavowed by Magellan's 
main investors, he left the company in July 1996 (New York Times, Jan. 12 1996;
Boston Herald, July 11 1996).}%
.
Such funds also have
much influence in {\it commercial} real estate; in contrast, their 
involvement in {\it residential} real estate has so far been smaller although this
situation may change in the next decades%
\qfoot{As an illustration one can mention a recent project of
North Western Mutual, one of
the largest American real estate investors. In March 2005, it launched
a \$ 350 million project in Virginia which combined 800 residential units
with 100,000 square meters of class A office space and a 256-room Marriott
hotel (New Port Daily Press, March 3 2005).}%
.
\qpar

Until recently, there was but little communication between stock and
real estate markets, but with the advent of financial globalization the 
boundaries between the two sectors tend to disappear. For instance,
on September 2004, Fidelity Investment opened a real estate branch
called Fidelity International Real Estate. For the time being the assets of
this fund remain fairly modest but there can be little doubt that this 
marks the start of a growth trend. The outcome 
of the present episode will show
how quickly the transformation occurs. If the amplitude of the peak
in Fig. 9 turns out to reach a level of  2.5 
(instead of 1.76 as in the graph) this will be evidence
that a major
metamorphose has already taken place in the Californian real estate 
market.

\vfill \eject

{\bf \large References}

\qparr
Bourgeois (R.) 1960: La crise \'egyptienne. 
Paris: Arthur Rousseau.

\qparr
Grison (C.) 1965: 1882, 1930 deux crises immobili\`eres dans l'agglom\'eration
parisienne. Revue d'Economie et de Droit Immobilier 16, 960-990.

\qparr
Homer (S.), Sylla (R) 1996: A history of interest rates.
New Brunswick (New Jersey): Rutgers University Press.

\qparr
Kaiser (R.W.) 1997: The long cycle in real estate. Journal of
Real Estate Research 14, 3, 233-257.

\qparr
Kaizoji (T.) 2005: Comparison of volatility distributions in the
periods of booms and stagnations: an empirical study on stock
price indices. 
(to be published).

\qparr
Maslov (S.), Roehner (B.M.) 2003: Does the price multiplier
effect also hold for stocks? 
International Journal of Modern Physics 14, 10, 1439-1451.

\qparr
Roehner (B.M.) 2000: Speculative trading: the price
multiplier effect. 
The European Physical Journal B 14, 395-399, 2000.

\qparr
Roehner (B.M.) 2001: Hidden collective factors in speculative
trading. 
Berlin: Springer.

\qparr
Roehner (B.M.) 2002: Patterns of speculation. 
Cambridge: Cambridge University Press.

\qparr
Roehner (B.M.) 2004a: Patterns of speculation in real estate and
stocks. in Hideki Takayasu, ed.: The application of econophysics, Proceedings
of the Second Nikkei Econophysics Symposium held in Tokyo, November 2002.
Tokyo: Springer.

\qparr
Roehner (B.M.) 2004b: Macro-players in stock markets.
Proceedings
of the Third Nikkei Econophysics Symposium held in Tokyo, November 2004.
To be published by Springer-Tokyo, Hideki Takayasu, editor.

\qparr
Wenzlick (A.) 1972: The Wenzlick 18.3-year cycle. 
Real Estate Analyst 1972-1973.

\qparr
Zhou (W.-X.), Sornette (D.) 2005: Is there a real-estate bubble in the
U.S.?
To appear in Physica A

\end{document}